\documentclass[conference, a4paper]{IEEEtran}
\usepackage{cite}
\ifCLASSINFOpdf
  \usepackage[pdftex]{graphicx}
\else
  \usepackage[dvips]{graphicx}
\fi
\usepackage[cmex10]{amsmath}
\usepackage{array}
\usepackage{theorem}
\usepackage[caption=false,font=footnotesize]{subfig}
\usepackage{fixltx2e}
\usepackage{url}

{\theorembodyfont{\rmfamily}
   }
{\theorembodyfont{\rmfamily}
   }
{\theorembodyfont{\rmfamily}
   }
{\theorembodyfont{\rmfamily}
   }

\begin{document}

\title{Increasing Physical Layer Security \\
through Scrambled Codes and ARQ}

\author{\IEEEauthorblockN{Marco Baldi, Marco Bianchi, Franco Chiaraluce,\\}
\IEEEauthorblockA{DIBET, Universit\`a Politecnica delle Marche,\\
Ancona, Italy\\
Email: \{m.baldi, m.bianchi, f.chiaraluce\}@univpm.it}}

\maketitle

\begin{abstract}
We develop the proposal of non-systematic channel codes on the AWGN wire-tap channel.
Such coding technique, based on scrambling, achieves high transmission security with 
a small degradation of the eavesdropper's channel with respect to the legitimate
receiver's channel.
In this paper, we show that, by implementing scrambling and descrambling on blocks of 
concatenated frames, rather than on single frames, the channel degradation needed
is further reduced.
The usage of concatenated scrambling allows to achieve security also when both
receivers experience the same channel quality.
However, in this case, the introduction of an ARQ protocol with authentication is needed.
\end{abstract}

\section{Introduction}
\label{sec:Intro}

A very simple model that is well suited to represent physical layer security schemes is the
wire-tap channel \cite{Wyner1975}, in which a transmitter 
(Alice) sends information to the legitimate receiver (Bob), but this is also received by the 
eavesdropper (Eve).
Alice can adopt whatever randomization, encoding and modulation scheme before transmitting
her message, and both Bob and Eve are perfectly aware of the transmission technique she uses; 
so, at least in principle, they are both able to recover the $1 \times k$ message vector
($\mathbf{u}$) from the $1 \times n$ codeword vector ($\mathbf{c}$).
However, the channel that separates Alice from Bob is generally different from that between Alice
and Eve. For this reason, the codeword received by Bob ($\mathbf{c}_\mathrm{B}$) is not equal to
that gathered by Eve ($\mathbf{c}_\mathrm{E}$).
So, after inverting the encoding map, the message obtained by Bob ($\mathbf{u}_\mathrm{B}$)
can differ from that recovered by Eve ($\mathbf{u}_\mathrm{E}$).


Under an information theoretic viewpoint, 
the wire-tap channel can be described through the secrecy capacity, defined as the
highest transmission rate at which Bob can achieve arbitrarily small error probability
while Eve has no information on the transmitted message.
In this study, we are interested in the Additive White Gaussian Noise (AWGN) wire-tap channel,
for which the secrecy capacity equals the difference between the two channel capacities \cite{Leung-Yan-Cheong1978}.
So, transmission security can be achieved only on condition that Bob's AWGN channel
has higher signal-to-noise ratio (SNR) with respect to Eve's AWGN channel. 
On the other hand, in \cite{Maurer1993} it was shown that physical layer security is attainable even when Bob's SNR is lower than Eve's SNR, but this requires the presence of a feedback channel between Alice and Bob (also accessible to Eve).
In this paper, we consider Automatic Repeat reQuest (ARQ) protocols \cite{Lin2004Book}, that 
exploit a very simple feedback mechanism, based on checking the integrity of all
frames and sending acknowledgment (or non-acknowledgment) messages from the receiver 
back to the transmitter.

The concept of secrecy capacity is based on limit values, and can be seen
as a bound when using real transmission techniques.
On the other hand, in real cases, it is interesting to estimate the channel 
degradation that is needed to reach a fixed level of physical layer security, 
though not achieving the secrecy capacity.
Furthermore, the secrecy capacity assumes that both receivers use optimal
decoders. Under a cryptographic viewpoint, the assumption of providing an attacker
with an optimal decoder would also be useful for estimating the minimum security
level of the system.
However, this is not necessary with the aim to make a comparison among different 
transmission techniques.
For this purpose, it is meaningful to suppose that both Bob and Eve use the same
decoding technique. If Eve would adopt a more efficient decoder, we can expect that 
the conclusion of the comparative analysis among different techniques does not change.

Based on these premises, our analysis is focused on a parameter named \textit{security gap}, 
that expresses the quality difference between Bob's and Eve's channels that is required
to achieve a sufficient level of physical layer security when both them implement
the same receiver.
An important target is to keep the security gap as small as possible, in such a way
as to achieve high physical layer security even with a small degradation of Eve's channel
with respect to Bob's one.

Some previous works have been devoted to the study of what transmission techniques are best
suited to reduce the security gap.
In particular, in \cite{Klinc2009}, the authors propose the usage of punctured codes,
by associating the secret bits to punctured bits.
They consider punctured LDPC codes and prove that such technique, 
for a fixed rate, is able to guarantee a considerable reduction in the security gap 
with respect to non-punctured (systematic) transmission.
Recently, we have proposed an alternative solution, based on non-systematic coding
implemented through scrambling of the information bits \cite{Baldi2010}. 
Such kind of coded transmission is able to achieve a strong reduction in the security gap, 
that becomes comparable (and even better) than that obtained through puncturing.

In this paper, we consider different coded transmission techniques over the AWGN
channel with Binary Phase Shift Keying (BPSK) and
 investigate how the security gap can be reduced by introducing concatenation within scrambling.
In this case, the information frames are grouped into blocks and scrambled/descrambled together,
so increasing the error propagation effect for the unauthorized receiver.
As a further contribution, we show that, when there is no gap between Bob's and Eve's channels,
security can still be achieved by resorting to an ARQ protocol with authentication by Bob.
In such case, only Bob can request for retransmission of a frame, that, however, is also
received by Eve.

The paper is organized as follows.
In Section \ref{sec:ScrambledCodes} we remind the main concepts and definitions of scrambled codes
for the wire-tap channel;
in Section \ref{sec:tErrorCorrecting} we study the case of $t$-error correcting codes under hard-decision
decoding;
in Section \ref{sec:LDPCCodes} we consider LDPC codes as an example of codes with soft-decision decoding;
in Section \ref{sec:arq} we describe the system adopting scrambled codes together with ARQ;
finally, Section \ref{sec:Conclusion} concludes the paper.

\section{Scrambled codes for the wire-tap channel}
\label{sec:ScrambledCodes}

Given the $k \times n$ generator matrix of an $(n, k)$-linear block code in systematic
form $\mathbf{G} = \left[ \mathbf{I} | \mathbf{C} \right]$, and a non-singular $k \times k$ 
binary scrambling matrix $\mathbf{S}$, 
Alice performs encoding as $\mathbf{c} = \mathbf{u \cdot S \cdot G}$, 
that is, by replacing the information vector $\mathbf{u}$ with its scrambled version 
$\mathbf{u' = u \cdot S}$ before applying the code.

If Bob's channel quality is sufficiently high,
Bob has a high probability to correct all errors, thus recovering, after descrambling, $\mathbf{u}_\mathrm{B} = \mathbf{u} = \mathbf{c}_l \cdot \mathbf{S}^{-1}$, where $\mathbf{c}_l$ is the vector containing the first $k$ bits of $\mathbf{c}$.
On the other hand, at the output of the descrambler, Eve has $\mathbf{u}_\mathrm{E} = \mathbf{u} + \mathbf{e}_l \cdot \mathbf{S}^{-1}$, where $\mathbf{e}_l$ is the left part of Eve's residual error vector $\mathbf{e} = \left[\mathbf{e}_l | \mathbf{e}_r \right]$.
So, if matrix $\mathbf{S}$ is suitably chosen, descrambling has the effect of propagating the residual errors after decoding.
A sufficient level of security is achieved when Bob's and Eve's energy per bit to noise power spectral density ratios (denoted by $\left. \frac{E_b}{N_0} \right|_\mathrm{B}$
and $\left. \frac{E_b}{N_0} \right|_\mathrm{E}$, respectively) are such that
Bob's bit error probability is lower than a given threshold, $P_e^\mathrm{B} \leq \overline{P_e^\mathrm{B}}\left(\left. \overline{\frac{E_b}{N_0}} \right|_\mathrm{B}\right)$, 
while Eve's one is greater than another threshold, $0.5 \geq P_e^\mathrm{E} \geq \overline{P_e^\mathrm{E}}\left(\left. \overline{\frac{E_b}{N_0}} \right|_\mathrm{E}\right)$.
The security gap is then defined, in dB, as $S_g = \left. \overline{\frac{E_b}{N_0}} \right|_\mathrm{B} - \left. \overline{\frac{E_b}{N_0}} \right|_\mathrm{E}$.

We obtain a first estimate of the scrambling effect by studying an ideal case we denote as \textit{perfect scrambling}: 
it models a scrambling technique that, in presence of one (or more) error(s), produces maximum uncertainty.
Under the hypothesis of perfect scrambling, a single residual bit error in the decoded word is sufficient to ensure 
that half of the bits are in error after descrambling.
In practice, perfect scrambling can always be approached by using $\mathbf{S}$ matrices with dense inverse
($\mathbf{S}^{-1}$), that is, with a density of $1$ symbols next to $0.5$.

The propagation effect of scrambling matrices on residual errors can be further
increased by implementing the scrambling and descrambling operations on blocks
of frames, rather than on single frames.
Let us suppose to collect $L$ consecutive information frames in a vector
$\overline{\mathbf{u}} = \left[\mathbf{u}_1 | \mathbf{u}_2 | \ldots | \mathbf{u}_L \right]$.
Scrambling can be directly applied on the whole $L$-frame block as:
\begin{equation}
\overline{\mathbf{u}'} = \overline{\mathbf{u}} \cdot \overline{\mathbf{S}},
\label{eq:ConcScramb}
\end{equation}
where $\overline{\mathbf{u}'}$ is the scrambled version of the $L$-frame block
and $\overline{\mathbf{S}}$ is a scrambling matrix with size $kL \times kL$.
After block scrambling, vector $\overline{\mathbf{u}'}$ is divided into its
components $\left[\mathbf{u}'_1 | \mathbf{u}'_2 | \ldots | \mathbf{u}'_L \right]$,
that are encoded and transmitted separately, as in the case without block scrambling.
Both Bob and Eve must collect their received frames into blocks
of $L$ frames before applying the block descrambling matrix $\overline{\mathbf{S}}^{-1}$.
After descrambling, Eve gets a block of $L$ information frames
$\overline{\mathbf{u}_\mathrm{E}} = \overline{\mathbf{u}} + \overline{\mathbf{e}_l} \cdot \overline{\mathbf{S}}^{-1}$,
where $\overline{\mathbf{e}_l}$ is the concatenation of
the error vectors affecting the $L$ information frames.

It is evident that, by using descrambling on blocks of frames,
its error propagation effect increases, since a single residual bit error
in one of the $L$ frames can be spread into all of them.
We extend the concept of perfect scrambling by defining that, under the 
hypothesis of a perfect block scrambler, a single bit error
in one of the $L$ frames ensures that, after descrambling: i) all the $L$ 
frames are in error and ii) half of the bits in each frame are erred.
We will see in the following that the condition of perfect scrambling can
be approached, even in such case, for practical choices of the system parameters.

\section{\textit{t}-Error Correcting Coding}
\label{sec:tErrorCorrecting}

Let us consider a hard-decision decoded $(n, k)$ linear block code
able to correct $t$ bit errors.
When such a coding scheme is adopted, the frame error probability,
under bounded-distance decoding, can be estimated as $P_f = \sum^n_{i=t+1} {n \choose i} P_0^i (1-P_0)^{n-i}$,
where $P_0$ is the channel bit error probability
taking into account the bandwidth expansion due to the code, 
i.e., $P_0 = 1/2 \cdot \mathrm{erfc}\left(\sqrt{E_b/N_0 \cdot k/n}\right)$.

When scrambling is performed over single frames, the frame error probability
is the same as for the unscrambled transmission; so, in presence of perfect scrambling, 
the bit error probability after descrambling is:
\begin{equation}
P_e^{PS} = \frac{1}{2}P_f = \frac{1}{2} \displaystyle\sum^n_{i=t+1} {n \choose i} P_0^i (1-P_0)^{n-i}.
\label{eq:tErrorCodingPS}
\end{equation}

In order to remove the hypothesis of perfect scrambling, we consider
a real descrambling matrix $\mathbf{S}^{-1}$ with column weight $w \leq k$
and denote: by $P_j$ the probability that a received $k$-bit vector contains $j$ errors
before descrambling; by $P_{i|j}$ the probability that exactly $i$
out of $j$ errors are selected by a weight $w$ column of $\mathbf{S}^{-1}$.
Under such assumptions, the bit error probability on each
received bit, after descrambling, can be calculated as follows:

\begin{equation}
P_e^{S} = \sum_{j=0}^k{P_j}\sum_{\begin{subarray}{c} i=1\\ i \ \mathrm{odd} \end{subarray}}^{\min\left(j,w\right)}{P_{i|j}}.
\label{eq:PeS}
\end{equation}
In \eqref{eq:PeS}, $P_j = {k \choose j}\sum_{i=t+1}^n{{n-k \choose i-j}P_0^i\left(1-P_0\right)^{n-i}}$ and $P_{i|j} = {j \choose i}{{k-j} \choose {w-i}}/{k \choose w}$.
We have observed in \cite{Baldi2010} that a density $w/k \approx 0.01$ is enough to approach the effect of perfect scrambling.

When scrambling is applied on blocks of $L$ concatenated frames, its effect can still be 
evaluated by considering the perfect scrambling model.
In this case, the frame and bit error probability after descrambling can be easily estimated as follows:
\begin{equation}
\left\{
\begin{array}{l}
P_f^{L\textrm{-}PS} = 1 - \left(1 - P_f \right)^L, \\
P_e^{L\textrm{-}PS} = \frac{1}{2}P_f^{L\textrm{-}PS}.
\end{array}
\right.
\label{eq:ConcPerfScramb}
\end{equation}

When real block scrambling matrices are instead adopted,
the bit error probability after block descrambling can be
estimated starting from the bit error probability with
single frame descrambling, given by \eqref{eq:PeS}.
Let us consider, for simplicity, that the $kL \times kL$ block
descrambling matrix has constant row and column weight $wL$, and it
is formed by $L \times L$ square blocks with size $k \times k$ and row/column
weight $w$.
Under such assumption,
each received bit after block descrambling can be seen as the 
sum of $L$ received bits after single frame descrambling; so,
its error probability can be estimated as:
\begin{equation}
P_e^{L\textrm{-}S} = \sum_{\begin{subarray}{c} i=1\\ i \ \mathrm{odd} \end{subarray}}^{L}{{L \choose i}\left(P_e^S\right)^i\left(1-P_e^S\right)^{L-i}}.
\label{eq:PeLS}
\end{equation}

\begin{figure}[t]
\begin{centering}
\includegraphics[width=83mm,keepaspectratio]{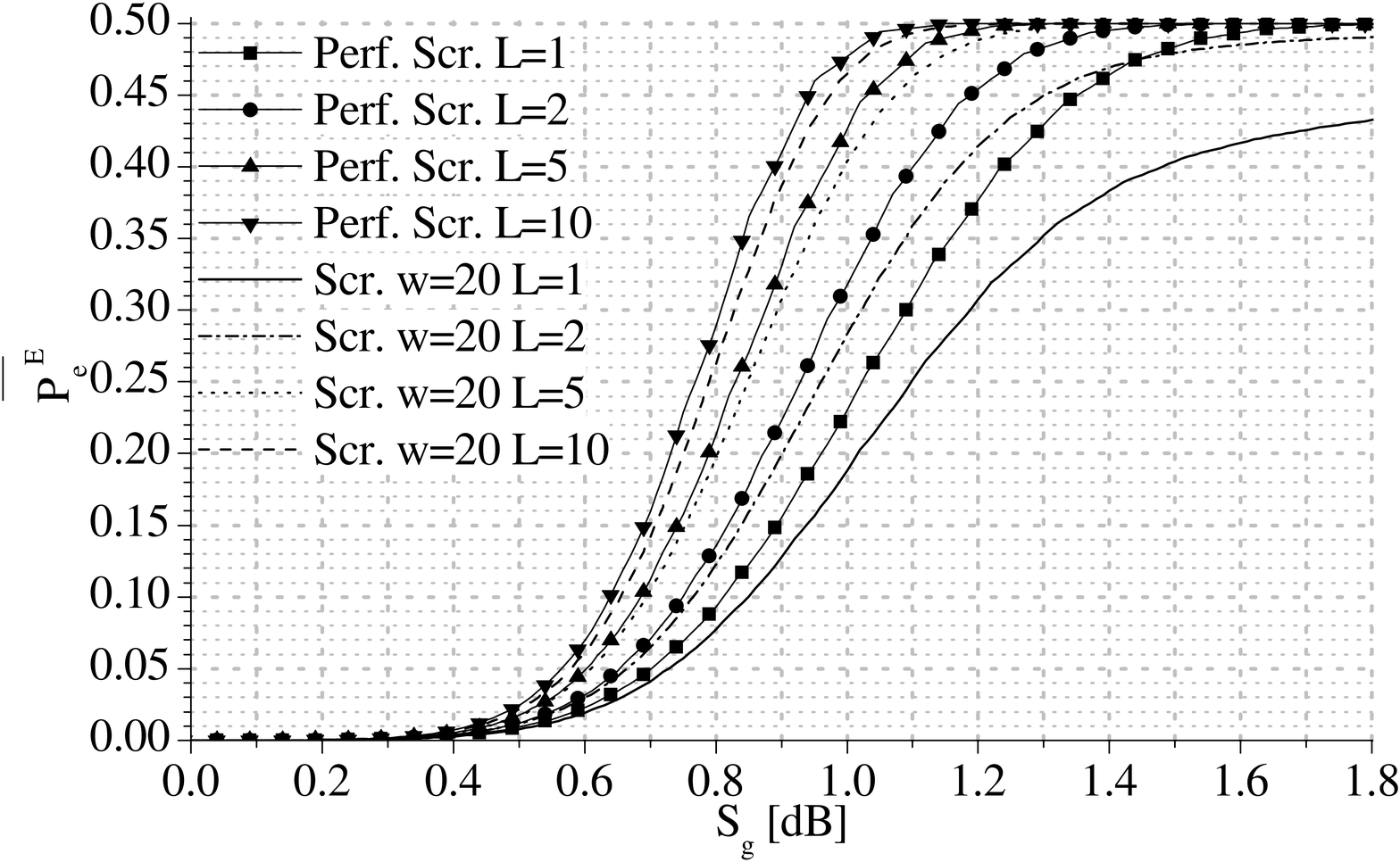}
\caption{Eve's bit error probability versus the security gap for $\overline{P_e^{\mathrm{B}}} = 10^{-5}$,
when using the $(2047,1354,69)$ BCH code with concatenated scrambling. \label{fig:tErrorCorrecting_Conc}}
\par\end{centering}
\end{figure}

In order to evaluate the effect of concatenated scrambling on the security gap,
we have considered a $(n=2047,k=1354,t=69)$ BCH code and $\overline{P_e^{\mathrm{B}}} = 10^{-5}$.
Fig. \ref{fig:tErrorCorrecting_Conc} shows the values of $\overline{P_e^{\mathrm{E}}}$, as a function 
of the security gap, when scrambling is performed over blocks of concatenated frames with different size ($L$).
As we observe from the figure, the introduction of concatenated scrambling reduces the
security gap needed to reach high values of $\overline{P_e^{\mathrm{E}}}$ with respect to the case 
without concatenation ($L=1$).
We also observe that the effect of a block descrambling matrix with row and column
weight $wL = 20L$, for $L>2$, approaches that of the corresponding perfect scrambler.
The convergence to perfect scrambling would also improve by increasing $w$.

\section{Non-Systematic LDPC codes}
\label{sec:LDPCCodes}

As an example of Soft-In Soft-Out modern error correcting schemes, we consider
LDPC codes, to which we apply the approach of non-systematic transmission based
on scrambling.
We consider an LDPC code with length $n = 2364$ and dimension $k = 1576$,
designed through the well-known Progressive Edge Growth (PEG) algorithm \cite{Hu2001PEG}.

In the case of scrambling performed over single frames, non-systematic LDPC coded transmission
has been obtained by adopting a dense $1576 \times 1576$ scrambling matrix, and we have
verified it practically reaches the effect of a perfect scrambler.
Also in this case, the security gap can be reduced by resorting to concatenated scrambling.
Starting from the $P_f$ values for the case with single frame scrambling, 
the effect of a perfect scrambler on blocks 
of $L$ concatenated frames can be estimated by means of Eqs. \eqref{eq:ConcPerfScramb}.
\begin{figure}[t]
\begin{centering}
\includegraphics[width=83mm,keepaspectratio]{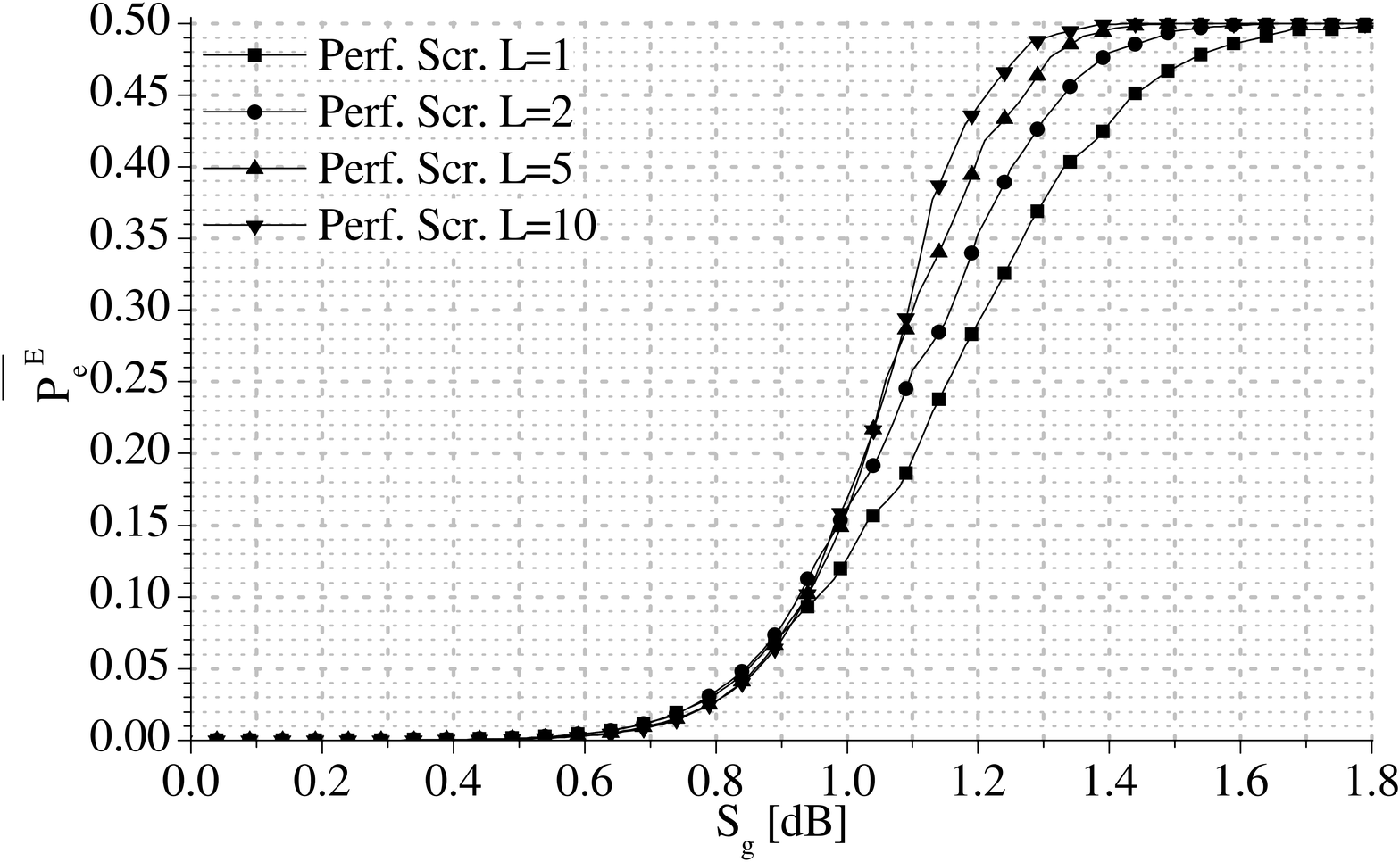}
\caption{Eve's bit error probability versus the security gap for $\overline{P_e^{\mathrm{B}}} = 10^{-5}$,
when using a $(2364,1576)$ LDPC code with concatenated scrambling. \label{fig:LDPC_Conc}}
\par\end{centering}
\end{figure}
Fig. \ref{fig:LDPC_Conc} shows the values of $\overline{P_e^{\mathrm{E}}}$ versus the security gap
for $\overline{P_e^{\mathrm{B}}} = 10^{-5}$ and different values of $L$. 
We have verified through numerical simulations that, by using real scrambling matrices
with a high density of $1$ symbols in their inverse, the effect of block scrambling is 
almost coincident with that predicted through the perfect scrambler approximation.

\section{Coded transmission with ARQ}
\label{sec:arq}

Coded transmission based on scrambling is able to strongly reduce the security gap
with respect to systematic transmission \cite{Baldi2010}.
However, physical layer security is still achievable only when Bob has a better channel
than Eve.
It is known from the literature that, when Bob's and Eve's channels have the same quality,
or Eve's channel is even better than Bob's, a feedback mechanism is needed
to achieve security \cite{Lai2008}, \cite{Harrison2010}.

We investigate such case by considering a very common feedback mechanism, based
on integrity checks and ARQ.
Obviously, the request for retransmissions must be allowed for Bob only; so, this
setting requires the existence of a form of authentication between Alice and
Bob.
However, retransmitted packets are also available to Eve, through her channel.

Physical layer security on channels with ARQ has been already studied and it
has been shown that a scheme based on ARQ can recover
security of the wired equivalent privacy (WEP) protocol for wireless local area
networks, otherwise vulnerable to total break attacks \cite{Omar2009}.
ARQ has also been recently used together with puncturing for physical layer
security: in \cite{Harrison2010}, the authors propose a scheme for using puncturing and
ARQ on the wire-tap channel with packet erasures in such a way that the loss of a single 
packet is sufficient to prevent the eavesdropper
from decoding the message.

In this paper, we use ARQ together with Forward Error Correcting (FEC) codes, so
we consider Hybrid ARQ (HARQ) schemes.
Several implementations are possible for HARQ schemes:
two main families are those using incremental redundancy \cite{Mandelbaum1974}
and soft-combining \cite{Holland2005}; both of them have been implemented in a number of variants.
When soft-decision maximum a posteriori probability (MAP) decoding algorithms are adopted, 
a common approach consists in using the reliability values obtained after each (failed) decoding attempt
as \textit{a priori} values for decoding the next replica of the same frame.
Other variants (as that proposed in \cite{Holland2005}) of such approach can be conceived that
are able to improve the performance.

However, this soft-combining strategy is only applicable with soft-decision decoding, while
we are interested in evaluating the performance of HARQ systems in less restrictive conditions.
For this reason, we consider a HARQ scheme that exploits the
simplest form of soft-combining: it consists in averaging the channel outputs, after
multiple transmissions of the same frame.
Such scheme is a valuable benchmark, since it can be implemented with
whatever family of codes (exploiting both hard and soft-decision decoding).
More complex strategies of soft-combining could provide some improvement in performance,
especially when the maximum number of allowed retransmissions is high (that is not the situation 
considered in this paper). In that case, Eve could get some advantage from adopting such 
more involved approaches. This will be the object of future work.


In the HARQ protocol we consider, Bob can exploit a number of transmissions $Q \leq Q_{\max}$ for decoding each frame
and Eve receives all retransmissions requested by Bob.
It should be noted that the integrity check mechanism, based on parity-checks, is exposed to
undetected errors, that is, transitions of the received codeword to near codewords.
In this case, integrity of the frame is erroneously verified. However, for long codes with
rather high rate and error correction capability, undetected errors are extremely rare
(as we have verified in our simulations).

According with the description above, a first solution is to try decoding only on the average of the
$Q$ received replicas of each frame. 
A more ``aggressive'' approach (that could be adopted by Eve) is to try all possible averages
of the $Q$ received replicas of each frame.
In this case, a further improvement in performance can be achieved (though at the cost of increased
complexity).
In general terms, starting from the curve of $P_f\left(E_b/N_0\right)$ for the considered
transmission scheme, the frame error probability when decoding is performed on the average 
of $Q \leq Q_{\max}$ independent transmissions can be modeled through an AWGN channel with noise
variance divided by $Q$; so, it becomes:
\begin{equation}
P_f^{(Q)}\left(E_b/N_0\right) = P_f\left(E_b/N_0 \cdot Q\right).
\label{eq:PfQ}
\end{equation}

In classical ARQ schemes, the frame error probability at each transmission of a given frame is 
independent of the others; so, the frame error probability after $Q$ transmissions is simply
the product of the $Q$ frame error probabilities at each transmission.
When using soft-combining, such hypothesis is no longer valid.
For example, in the soft-combining strategy we adopt, since the second transmission of a frame
is requested only upon a decoding failure on the first transmission, 
the decision variable is no longer Gaussian and \eqref{eq:PfQ} should be modified.
However, as we have verified through simulations, the impact of such correction is negligible,
at least for our choice of the system parameters; so, we adopt $P_f^{(Q)}$, expressed by \eqref{eq:PfQ}, as
an estimate of the frame error probability at the $Q$-th transmission of the considered HARQ protocol.

In the system model we adopt, Bob is always able to request retransmission of a frame, when
needed (i.e., after a decoding failure); so, the probability that he receives $Q \geq 1$ transmissions 
of a frame coincides with his frame error probability after $Q-1$ transmissions, that is:
\begin{equation}
\left. P_R^{(Q)} \right|_{\mathrm{B}} = \prod_{i=0}^{Q-1} P_f^{(i)},
\label{eq:PRQB}
\end{equation}
with $P_f^{(0)} = 1$.
On the contrary, when Eve, after a decoding failure, would need a frame retransmission,
she is able to receive it only on condition that Bob has failed to decode the same frame.
Then, the probability that she receives $Q \geq 1$ (useful) transmissions of a frame is:
\begin{equation}
\left. P_R^{(Q)} \right|_{\mathrm{E}} = \prod_{i=0}^{Q-1} \left[ P_f^{(i)} \right]^2.
\label{eq:PRQE}
\end{equation}
Based on \eqref{eq:PRQB} and \eqref{eq:PRQE}, Bob's and Eve's frame error probability is:
\begin{equation}
\left. P_f^{\mathrm{B}/\mathrm{E}} \right|_{\mathrm{ARQ}} = 1 - \displaystyle\sum_{i=1}^{Q_{\max}} \left. P_R^{(i)} \right|_{\mathrm{B}/\mathrm{E}} \cdot \left(1 - P_f^{(i)} \right).
\label{eq:PfARQ}
\end{equation}

In the following subsection, we present some results on the usage of the HARQ scheme, here
described, with the $t$-error correcting codes and LDPC codes considered in the previous
sections.

\subsection{ARQ with $t$-Error Correcting codes}
\label{subsec:arq-bch}

\begin{figure}[t]
\begin{centering}
\includegraphics[width=83mm,keepaspectratio]{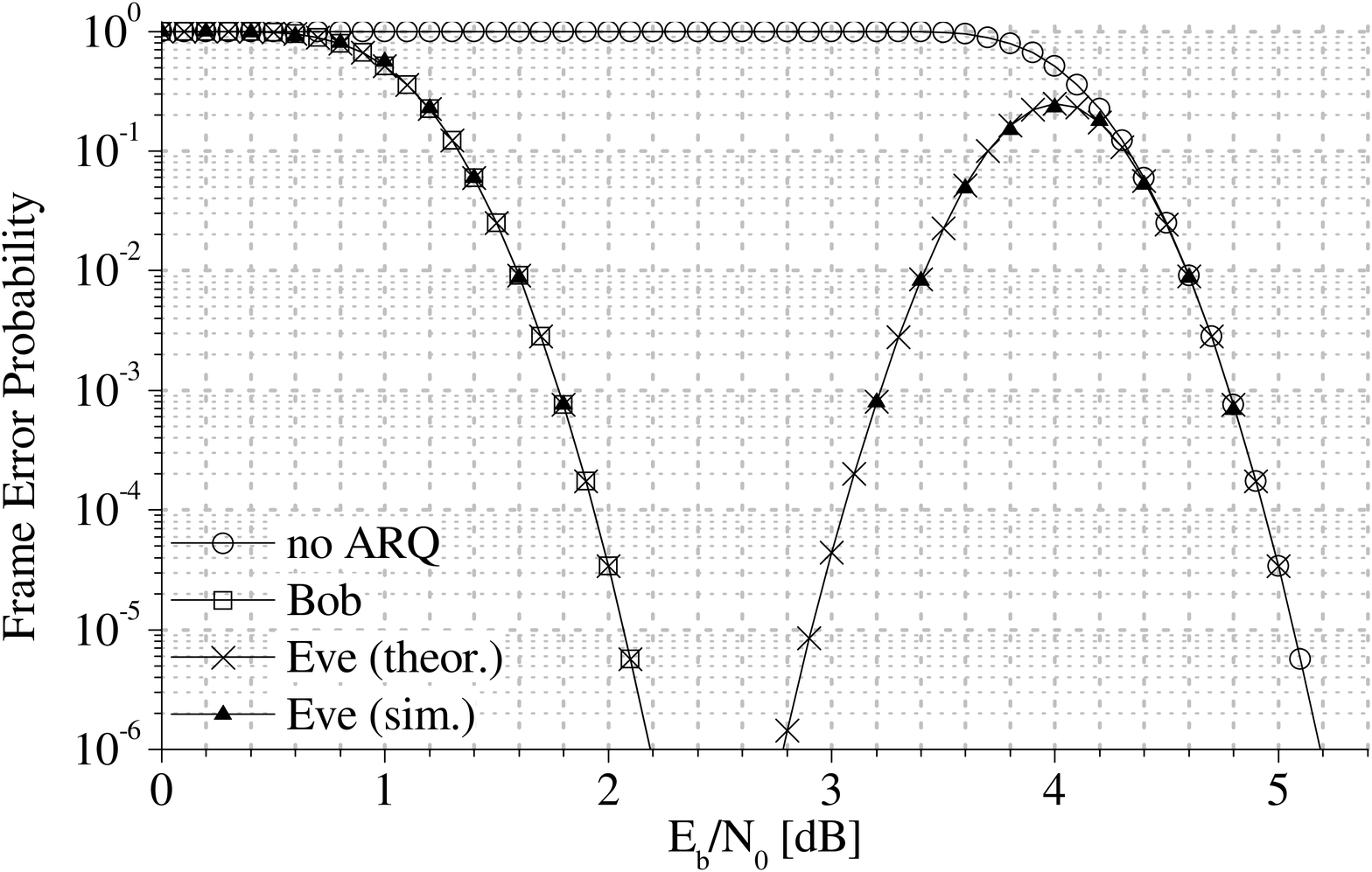}
\caption{Frame Error Probability versus SNR for the $(2047,1354,69)$ BCH code with soft-combining HARQ ($Q_{\max} = 2$). \label{fig:ARQ-BCH}}
\par\end{centering}
\end{figure}

Fig. \ref{fig:ARQ-BCH} shows the frame error probability curves, for the $(2047,1354,69)$ BCH code, obtained
through \eqref{eq:PfARQ} and by numerical simulations. The maximum number of transmissions has been
set $Q_{\max} = 2$.
As we observe from the figure, the analytical model fits very well the simulated data.
From the frame error probability curves we can deduce the bit error probability performance in presence
of scrambling, by using the simplifying hypothesis of perfect scrambling.

From the figure we notice that Bob's and Eve's performance can be divided into three
regions of SNR values: 
\begin{enumerate}
\item At low SNRs, Bob's channel quality is poor ($P_f^{\mathrm{B}} = 1$). In this region,
Bob requests retransmission of all frames, and Eve can benefit from this fact. So, Eve's
performance is overlaid with Bob's one, and their error probability curves have a gain of $3$ dB
(due to $Q = Q_{\max} = 2$) with respect to the case without ARQ, as expected.
\item When the channel quality is high enough that Bob requests few retransmissions ($P_f^{\mathrm{B}} < 1$), 
Eve misses most of the retransmissions she would need. Moreover, the channel quality is still not
sufficient to allow correct decoding without ARQ, so her performance deteriorates.
Bob's error probability curve continues to decrease towards very small values, while Eve's one grows.
\item For large SNR values, the channel quality becomes adequate to allow decoding with a
single transmission of each frame, and Eve's performance tends to that of the system without ARQ.
\end{enumerate}

An important observation (that also justifies the interest for the frame error probability, in place 
of the bit error probability, as in the previous figures) is that, without resorting to concatenated scrambling, HARQ
is not able to achieve a sufficient level of physical layer security. In fact, in the region where Bob achieves
low frame error probabilities, Eve's frame error probability is always too low to guarantee security.
However, by using concatenated scrambling, the security condition can be recovered.

For example, by means of Eqs. \eqref{eq:ConcPerfScramb}, we can easily estimate that, by using
a concatenated scrambler with $L = 20$, the region of $P_f^{\mathrm{E}} \geq 10^{-1}$ (around $4$ dB)
translates into $P_f^{\mathrm{E}} > 0.87$.
In the same region, Bob's frame error probability is surely $P_f^{\mathrm{B}} < 10^{-6}$, and it becomes
$P_f^{\mathrm{B}} < 2 \cdot 10^{-5}$ after the application of the concatenated scrambler.
So, under the perfect scrambler assumption, it is $P_e^{\mathrm{E}} > 0.4$
and $P_e^{\mathrm{B}} < 1 \cdot 10^{-5}$, that restores the desired security level.

\subsection{ARQ with LDPC codes}
\label{subsec:arq-ldpc}

\begin{figure}[t]
\begin{centering}
\includegraphics[width=83mm,keepaspectratio]{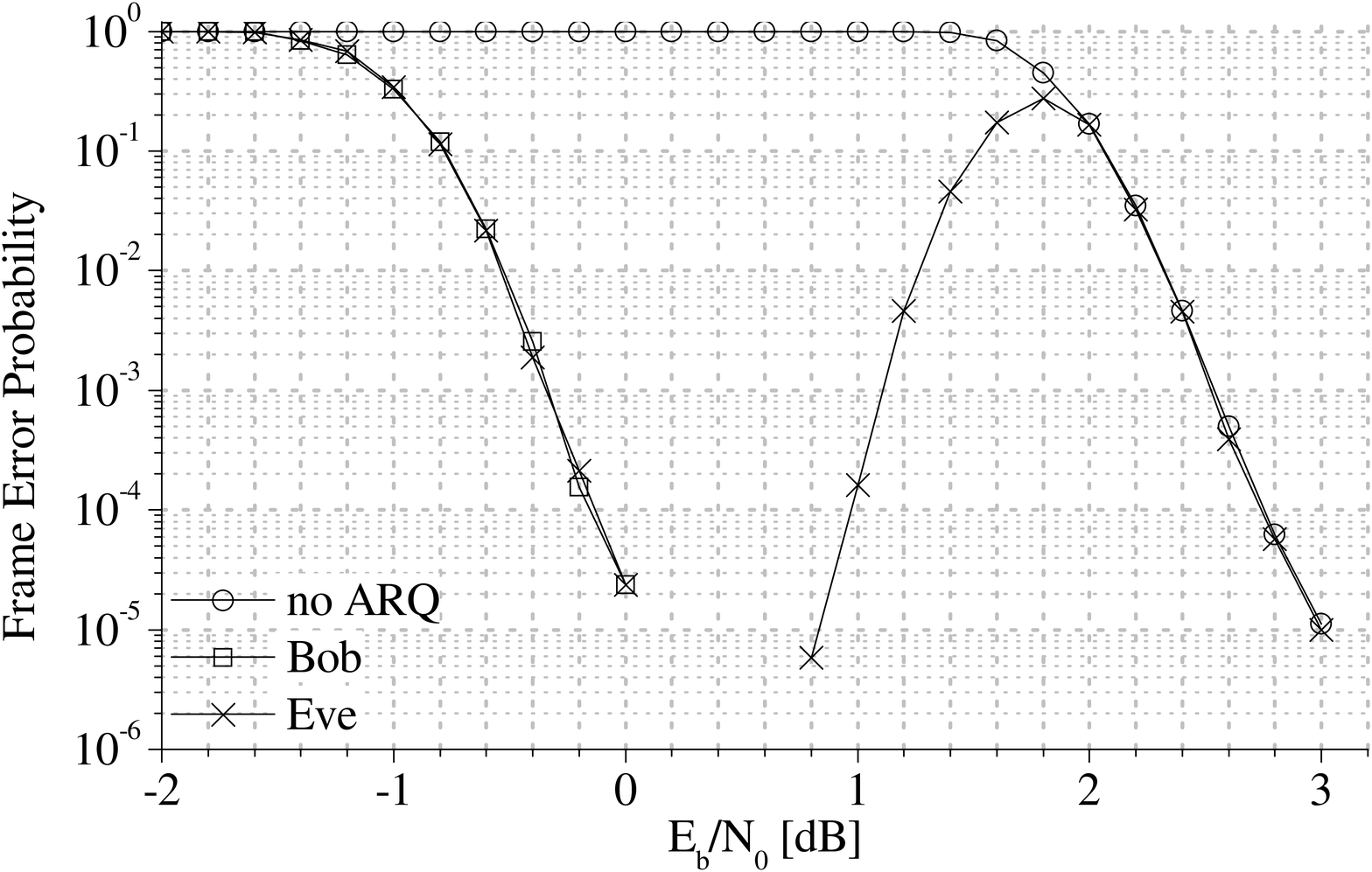}
\caption{Frame Error Probability versus SNR for the $(2364, 1576)$ LDPC code with soft-combining HARQ ($Q_{\max} = 2$). \label{fig:ARQ-LDPC}}
\par\end{centering}
\end{figure}

A very similar situation is observed for ARQ with LDPC codes, whose frame error probability curves
have been estimated through numerical simulations and are reported in Fig. \ref{fig:ARQ-LDPC}.
From the figure we notice that Bob's and Eve's error probability curves have the same
three-zone behavior observed for hard-decision decoded BCH codes.

Also in this case, we observe the presence of a region (around $1.8$ dB) where Eve's frame
error probability is $\geq 10^{-1}$ while Bob's one (according to the trend of the simulated curve) 
will become $< 10^{-6}$; so, the application of a concatenated scrambler with $L = 20$ still 
allows to achieve the desired physical layer security.

\section{Conclusion}
\label{sec:Conclusion}

We have continued the study of codes with scrambling for physical layer security over the AWGN wire-tap channel.
Our results show that the implementation of scrambling and descrambling on blocks of concatenated frames,
rather than on single frames, can further reduce the security gap.

By using concatenated scrambling in conjunction with a HARQ protocol with soft-combining, a sufficient level
of physical layer security can also be ensured when Bob's and Eve's channels have the same quality.
The same target could be reached even when Eve has a better channel than Bob; this aspect will
be investigated in future works.

\newcommand{\BIBdecl}{\setlength{\itemsep}{0.01\baselineskip}}
\bibliographystyle{IEEEtran}
\bibliography{Archive}

\end{document}